\documentclass{article}

\usepackage{arxiv}
\usepackage{graphicx}
\usepackage{amssymb,amsthm,amsmath}
\usepackage{xcolor,paralist,hyperref,titlesec,fancyhdr,etoolbox}
\usepackage{enumitem}
\usepackage{tikz}
\usepackage{cite}
\usepackage{amsmath}
\usepackage{amsfonts}
\usepackage{nicefrac}
\usepackage{microtype}
\usepackage{lipsum}
\usepackage{caption}  
\usepackage{tabularx}
\usepackage{booktabs}
\usepackage{siunitx}
\usepackage{makecell}
\usepackage{cellspace}
\usepackage{float}
\usepackage{indentfirst,csquotes}
\usepackage{url}

\title{RoNo: A novel way in generating reconfigurable on-chip nonlinear activation functions}

\author{Zili~Cai, Tian~Zhang*, Jian~Dai, Zheng~Wang, Kun~Xu \\
        \textit{State Key Laboratory of Information Photonics and Optical Communications} \\
        \textit{Beijing University of Posts and Telecommunications} \\
        \textit{Beijing, China} \\
        \url{ztian@bupt.edu.cn}
        }
\date{}

\begin{document}
\maketitle
\begin{abstract}
Due to the limitations of Moore's Law and the increasing demand of computing, optical neural network (ONNs) are gradually coming to the stage as an alternative to electrical neural networks. The control of nonlinear activation functions in optical environments, as an important component of neural networks, has always been a challenge. In this work, firstly, we use inverse design tools to design a optical patterned area in silicon-carbide-on-insulator. This patterned area could generate two different nonlinear responses of the amplitude. Secondly, the patterned region is integrated with a control network to form a reconfigurable on-chip nonlinear activation function generator for wave-based analog computing. Experiment shows that neural network that uses such a system as an activation function performs well in the MNIST handwritten dataset and CIFAR-10, respectively. Compared to previous works, we propose a novel approach to generate on-chip reconfigurable activation functions in optical neural networks, which achieves compact footprint and enables high-quality activation function generation. 
\end{abstract}

\section{Introduction}
In the past few decades, the world had witnessed the development of the neural network with the application of deep learning \cite{chen2025deep}. In many fields, artificial neural networks play a huge role, such as genetic engineering \cite{deeplearning_genomic}, materials science \cite{deeplearning_material}, image classification \cite{deeplearning_image}, and design of photonic structure \cite{deeplearning_photonic1}, \cite{deeplearning_photonic2}, \cite{deeplearning_photonic3}, \cite{deeplearning_photonic4}. However, with the increase of computing demand and data volume, traditional electrical neural network are faced with many problems and researchers have begun to find a new way to establish neural network with faster computational speed and lower consumption. Optical neural networks (ONNs) \cite{ONN1}, \cite{ONN2} have attracted more and more researchers' attention. Comparing with electronic neural networks, the integrated ONNs take an advangage of high computation rate and low energy loss. Nonlinear activation function (NAF), as an important part of ONNs, plays a crucial role in accelerating training process and enhancing the accuracy \cite{NAF1}, \cite{NAF2}. Generally, like electronic neural networks, ONNs with different architectures need different types of NAFs. Currently, it is mature to implement NLAFs using electrical means, but still is in infant stage using optical means. 

A growing number of efforts have been made in generating NLAFs \cite{zili_cai_inverse_design}. At the early stage of ONNs, electro-optic architecture has been reported to realize optical-to-optical activation functions \cite{electric_optical_NAF1}, \cite{electric_optical_NAF2}. This scheme offers high reconfigurability but brings a relatively large time delay because of the opto-electric conversion. To accelerate the operation speed of NLAFs, many all-optical schemes has been proposed. By changing the absorption for optical power or changing the resonant state, saturated absorber \cite{saturable_absorption_NAF1}, \cite{saturable_absorption_NAF2} and phase change materials combining with micro-ring resonators \cite{PCM_MRR_NAF} are both widely used in generating NAFs. Furthermore, by using nonlinear germanium-silicon photodiodes, silicon-integrated and monitorable optical neurons are realized \cite{germanium-silicon_NAF}. These strategies can realize quick-response NLAFs but only a fixed type of NLAF can be realized. It has been proved that the performance of neural networks \cite{NAF2}, \cite{NAF_comparison} can be significantly influenced by the type of NLAFs. Thus, realizing various types of NLAFs in optical circuits is of significance to ONNs, especially ONNs with complex architecture.

One approach to achieve various types of NLAFs is based on Mach-Zehnder interferometer (MZI) loaded with phase shift electrodes \cite{HuangChaoRan_reconfigurable}. Owing to the free-carrier dispersion (FCD) effect in the MRR cavity, there is a strong nonlinear phase shift response to optical power. Four types of NLAF can be achieved through changing the wavelength detuning between the input signal and the MRR resonance. Inverse design methods, mainly including evolutionary algorithm based methods \cite{inverse_design_NSGA}, machine-learning-based methods \cite{inverse_design_deeplearning1}, \cite{inverse_design_deeplearning2} and topology optimization methods \cite{adjoint_method_and_inverse_design_for_nonlinear_nanophotonic_devices}, have extensively applied in nanophotonics owing to its outstanding ability of designing devices or systems without the need of establishing theoretical models. An amount of nonintuitive and compact devices or systems have been designed with superior performance. Recently, a reconfigurable nonlinear activation element has been reported by combing an inverse designed nonlinear structure and tunable couplers \cite{Reconfigurable_nonlinear_optical_element_using_tunable_couplers_and_inverse-designed_structure}. However, more couplers are needed to make the obtained function closer to the target nonlinear function, making this element not compact enough. Besides, the nonlinear activation function is formed by the connection of several signals from nonlinear signal dividers, introducing several discontinuity points in the nonlinear activation function. Most of ONNs use gradient-based training method. This discontinuity points will make the ONN tend to experience oscillations and slow convergence, deteriorating the performance of ONNs and limiting its practical applications on large-scale networks.

In this paper, we present a reconfigurable on-chip NAF generator (RoNo) comprising a patterned area (PA) and three phase shifters. The passive PA component, with a compact footprint of $7\times 7 \mu m^2$, inherently generates two distinct nonlinear responses through its structural design. By coherently synthesizing these responses via phase-shifter control, RoNo enables dynamic reconfiguration of customized nonlinear activations without requiring structural modifications. Three different NAF types and two parametric modified ones were experimentally demonstrated through this platform. To evaluate neural network adaptability, we implement these NAFs in two benchmark neural networks: a basic CNN architecture achieving $96.34\%$ accuracy on MNIST classification and a modified ResNet-18 attaining $84.11\%$ accuracy on CIFAR-10 recognition, verifying functional versatility across complexity levels. The experimental results show that our NAFs realized by RoNo enable high accuracy in different kinds of tasks. To further validate the generality of RoNo, we benchmark its performance against conventional electronic activation functions (e.g., Softplus, Swish and Tanh) under same network architectures and training protocols. Remarkably, RoNo-generated NAFs achieve comparable accuracy while operating passively without external power injection, demonstrating their inherent energy efficiency. This characteristic is particularly advantageous for ONNs requiring cascaded nonlinear operations and these results position RoNo as a universal building block for scalable, energy-efficient and effective optical neural processors.

\section{Device design process}
\subsection{Configuration of RoNo structure}
\begin{figure*}[h!]
    \centering
    \includegraphics[width=0.8\textwidth]{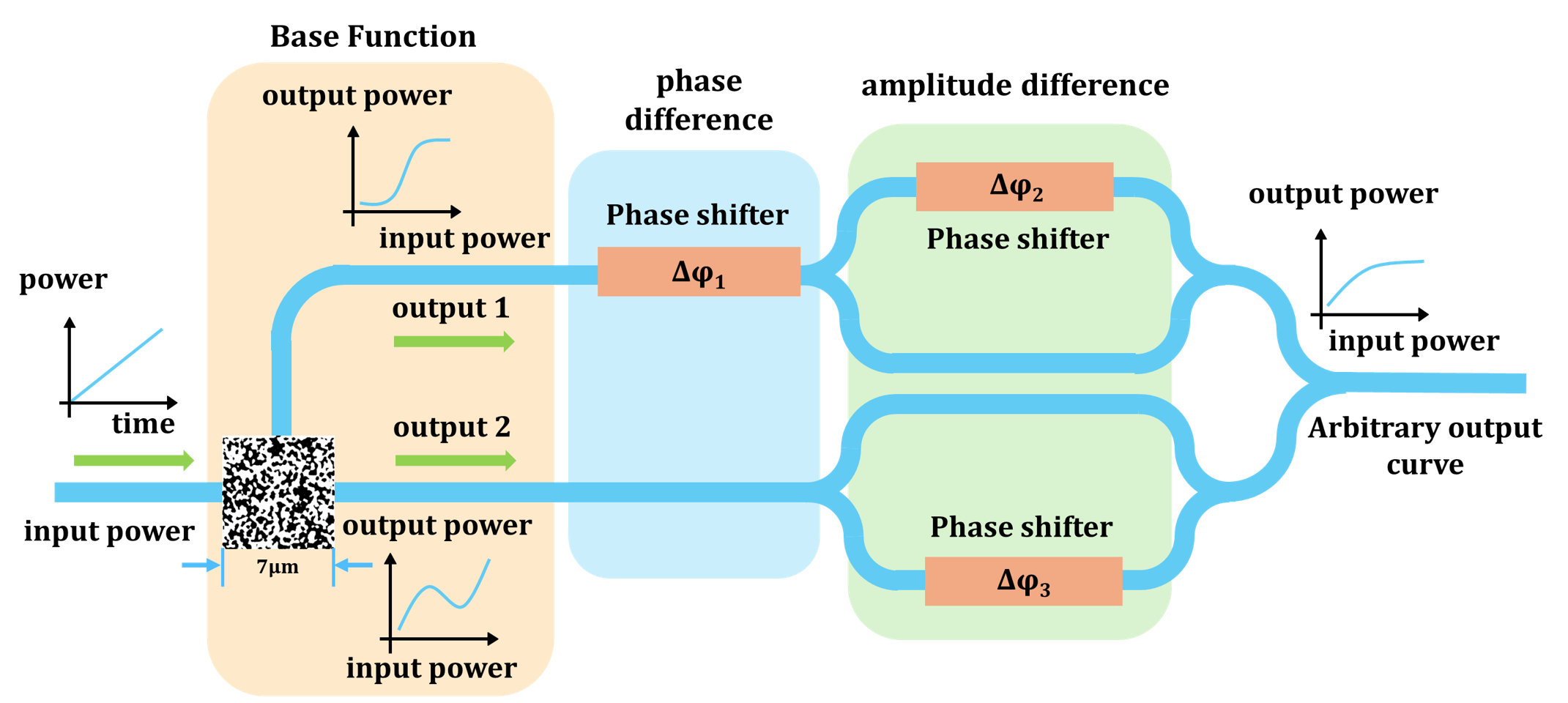}
    \caption{Schematic of RoNo}
    \label{fig:1}
\end{figure*}

In this work, we focus on the generation of on-chip arbitrary reconfigurable NAF in all-optical environment. As is shown in \autoref{fig:1}, the whole device consists of a patterned area (PA) and three phased shifters, corresponding to three parts of the RoNo.

In the base function section, there is an optimized compact PA ($7\times 7\mu m^2$) that generates the basic two nonlinear responses. The operating wavelength of the RoNo is 1.55$\mu m$. Fabricated in silicon-carbide (SiC) with its outstanding three-order nonlinear coefficient ($n_2=2.7\times10^{-14}cm^2/W$) and a relatively high relative dielectric constant (6.56) \cite{Measurement_of_the_Kerr_Nonlinear_Refractive_Index_and_its_Variation_Among_4H-SiC_Wafers}, which enables a strong Kerr nonlinear effect at a relatively low energy threshold (around $25mW/\mu m$). There are three ports of the PA, including an incident port and two output port. The incident light is split within the PA according to a predesigned nonlinear ratio and subsequently recombined in the following two sections. To enable the synthesis of more complex NAFs, the nonlinear ratio here requires customized design. Through a multi-object inverse design accelerated by adjoint method, we engineer the coupling region to simultaneously generate two nonlinear responses with different nonlinear characteristics. This ensures the subsequent MZI network can synthesize almost arbitrary NAFs through coherent superposition of the base functions. The waveguide is designed as 300nm in width which allows the propagation of the fundamental transverse electric (TE) mode. Here, the PA in our RoNo is designed and simulated using a 2D frequency-domain finite-difference (FDFD) method (adopting the open source program ANGLER \cite{adjoint_method_and_inverse_design_for_nonlinear_nanophotonic_devices}), the multi-object optimizing method and the synthesis process is calculated with python.
\subsection{System Design Process}
\begin{figure*}[h!]
    \centering
    \includegraphics[width=0.8\textwidth]{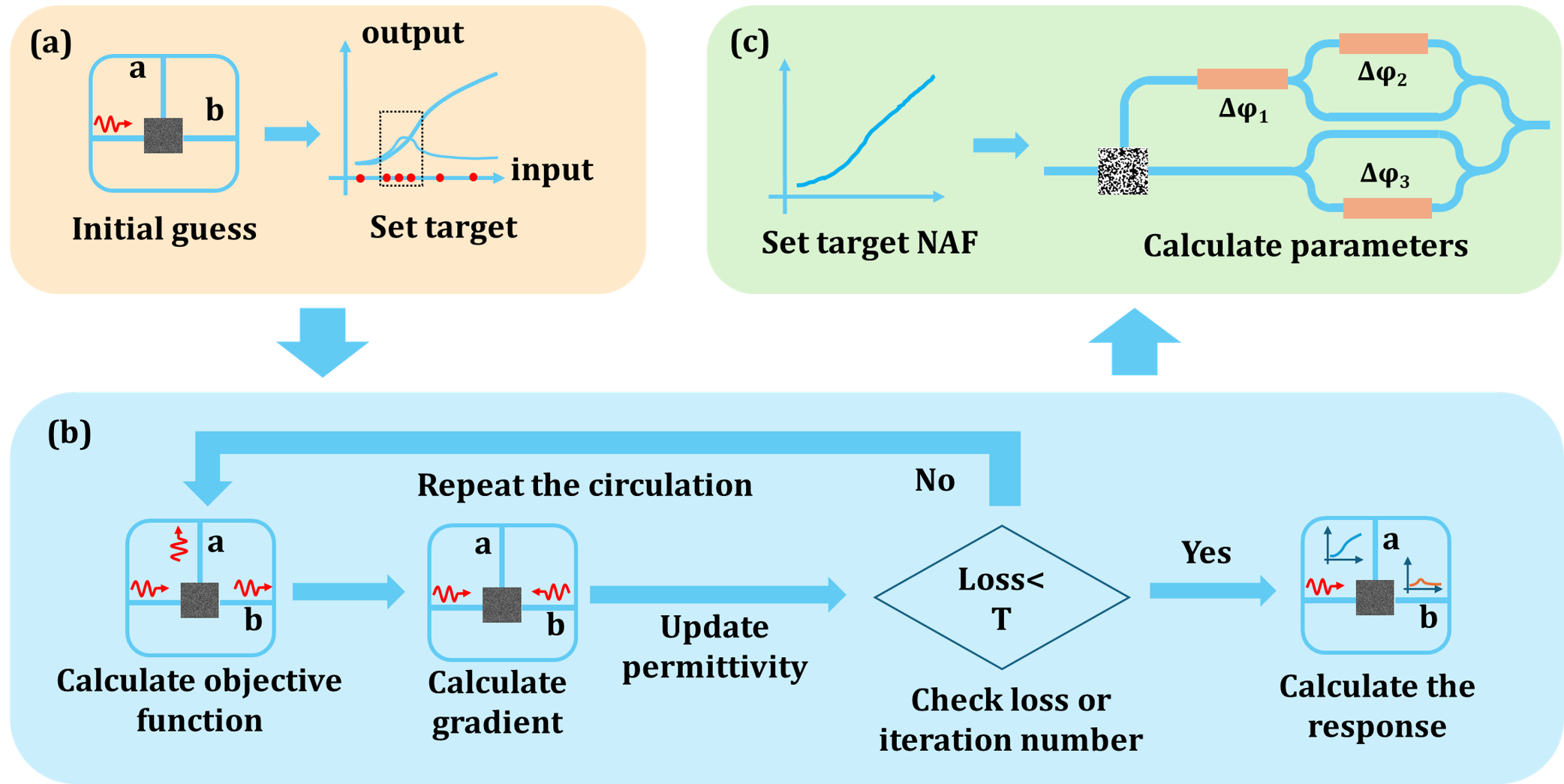}
    \caption{Flowsheet of the design process (a) initialization and set dual optimization target (b) optimization detail of PA (c) synthesis of customized NAFs}
    \label{fig:2}
\end{figure*}

The optimizing process of our RoNo is shown in \autoref{fig:2}. The whole process is split into three parts, including initialization optimization and synthesis.

(1) Initialize optimization area of PA. Firstly, we divide the design region into multiple grid cells, where the value in each grid represents an idealized and continuously variable relative dielectric constant. To reduce the impact of different random initializations on optimization results and accelerate the optimization, all grid values are set to half of the relative dielectric constant of SiC ($3.277$) \cite{Initial_value_for_inverse_design}. During the optimization, the relative dielectric constant is set to dependent on the intensity \cite{adjoint_method_and_inverse_design_for_nonlinear_nanophotonic_devices}, which is proportional to the square of the electric field strength:
\begin{equation}
    \label{eq:1}
    \tilde{\varepsilon}_r(r) = \varepsilon_r(r) + 3\omega_0^2 \varepsilon_0 \chi^{(3)}(r) I(r)
\end{equation}
where $\tilde{\varepsilon}_r(r)$ and $\varepsilon_r(r)$ represent the relative dielectric constants under Kerr nonlinear regime and the linear regime, repectively. The $\chi^{(3)}$ represent the real part of the third-order nonlinear polarizability. $I(r)$ represents the intensity.

(2) Set the optimization target. To enable more complex synthesis in the subsequent stage of RoNo, two distinct types of nonlinear responses need to be generated at the dual output ports and incorporated into the objective function during optimization. To quantify the statistical dependence between the energies of the two output ports ($E_{out1}, E_{out2}$) under incident light excitation, this work employs a kernel density estimation (KDE)-based mutual information (MI) approach. The MI can be calculated as below:
\begin{equation}
    I(e_t; e_r) = \iint p(e_t, e_r) \log\left( \frac{p(e_t, e_r)}{p(e_t)p(e_r)} \right) \mathrm{d}e_t \mathrm{d}e_r
    \label{eq:2}
\end{equation}
Where $e_t$ and $e_r$ represent the output power of the top and right port, respectively. The joint density $p(et, er)$ and the marginal densities $p(et), p(er)$ are estimated via Gaussian KDE, where the bandwidth is calculated using Scott’s rule:
\begin{equation}
    h = n^{-\frac{1}{d + 4}}\sigma
    \label{eq:3}
\end{equation}
Where $n$ is the number of the samples, $d$ denotes the data dimension and $\sigma$ represents the scaling factor of the covariance matrix. Here, we employ a modified Sigmoid function for the top port, while the lower port adopts its complementary function mirrored about the energy conservation curve ($y=x$). Both functions are as follows:
\begin{align}
    f_{\text{t}}(e_{\text{in}}) 
        &= \frac{a}{1 + e^{-be_{\text{in}} + c}} \label{eq:4}\\
    f_{\text{r}}(e_{\text{in}}) 
        &= 1 - f_{\text{t}}(e_{\text{in}}) \nonumber \\
        &= 1 - \frac{a}{1 + e^{-be_{\text{in}} + c}} \label{eq:5}
\end{align}
Where $f_t$ and $f_r$ represent the target transmission curve at the top and right port, respectively. By applying discrete sampling points of incident power as employed in this work to compute the target outputs and calculating the MI, the MI between these two objective functions is approximately 0.143, which aligns with expectations.

(3) Calculate the objective funtcion. The objective function of the inverse design is described below:
\begin{equation}
    F_{\text{total}} = -\sum_{i=1}^{n} \left( |e_{i,\text{t}} - f_{\text{t}}(e_{i,\text{t}})| + |e_{i,\text{r}} - f_{\text{r}}(e_{i,\text{r}})| \right)
    \label{eq:6}
\end{equation}
Where $n$ is the number of the samples. This function consists of two components: the deviations between the actual output power and the target output power at both ports, with data summed across all sampling points. As the objective function increases, the discrepancy between actual and target outputs decreases.

(4) Calculate the global gradient. With the help of adjoint method, we only need to calculate the forward propagated field and the adjoint field in the PA and then get the global gradient \cite{adjoint_method_and_inverse_design_for_nonlinear_nanophotonic_devices}. Since this is an equal-weight multi-objective optimization, the gradient of the total objective function can be calculated by summing the gradients of individual sample objective functions.

(5) Update the permittivity of PA and estimate the iteration condition. We use the limited-memory Broyden–Fletcher–Goldfarb–Shanno (l-BFGS) \cite{Gauss-Newton_and_full_Newton_methods_in_frequency-space_seismic_waveform_inversion} to minimize the objective function. The maximum iteration number of optimization is set as 330 to balance the optimization time and final performance. Optimization stops when the iteration number exceeds the maximum or the real reduction of the objective function drops below a threshold ($10^{-5}$ here).

(6) Repeat step (3) to step (5) until the optimization stops.

(7) Calculate the response of PA at port a and port b by sweeping the incident power from 0 to 0.05 $W/\mu m$ with 100 samples.

(8) Generate reconfigurable NAFs through changing the phase shifters of RoNo. 
\section{Experiment results and nonlinear synthesis}
\subsection{Transmission curve of PA}
\begin{figure*}[h!]
    \centering
    \includegraphics[width=0.8\textwidth]{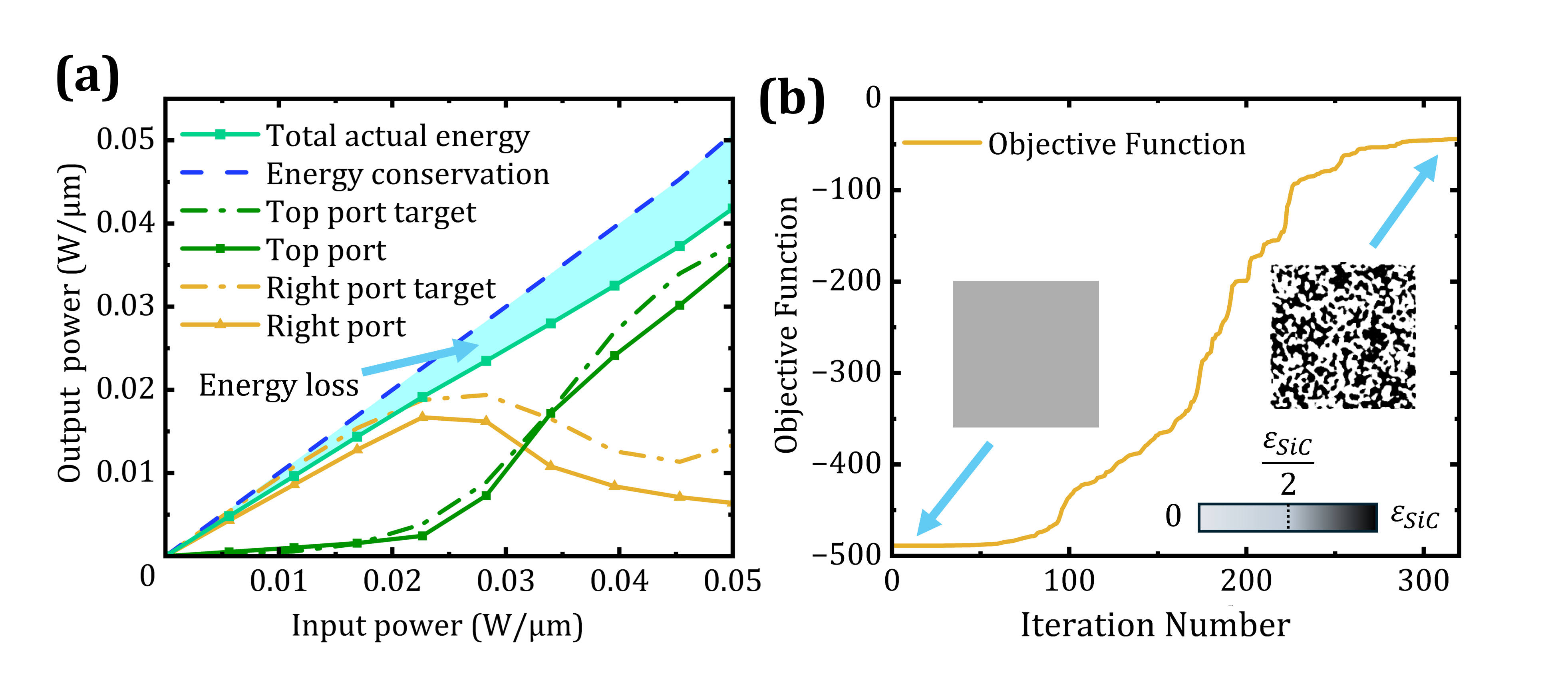}
    \caption{Evaluation of the PA and the design process (a) optimization performance and nonlinear responses of PA at the top port and the right port (b) objective function and the initial and final material distribution of PA}
    \label{fig:3}
\end{figure*}

The transmission curves for each port of the optimized PA are shown in \autoref{fig:3} (a). The measured transmissions at the upper and right ports are represented by green and yellow solid curves, respectively, with corresponding optimization targets shown as dashed lines. The blue dashed line represents the total transmission curve of the two output ports under lossless conditions, while the actual total transmission curve (light green solid line) deviates due to energy loss, with the region between them indicating the energy dissipation. \autoref{fig:3} (b) illustrates the objective function versus the number of optimization iterations. To achieve enhanced precision, the relative permittivity of all grid cells in the PA are defined as continuously variable values within the range (0, $\varepsilon_{SiC}$], as shown in the bottom color bar, which are subsequently transformed into the physical material distribution via a projection method upon optimization termination. During initialization, an idealized material distribution is employed, where all grid cells are assigned a value corresponding to half $\varepsilon_{SiC}$. The figure reveals that the optimization process converges after approximately 260 iterations, achieving an objective function value of -44.16 (about 9\% of the initial value), which demonstrates the excellent performance of our method. The transmission curves of the two output ports of the PA obtained from experiments were sampled, and the actual MI was calculated to be approximately 0.197, which aligns well with our optimization expectations.
\subsection{Synthesis of NAFs}
\begin{figure*}[h!]
    \centering
    \includegraphics[width=1\textwidth]{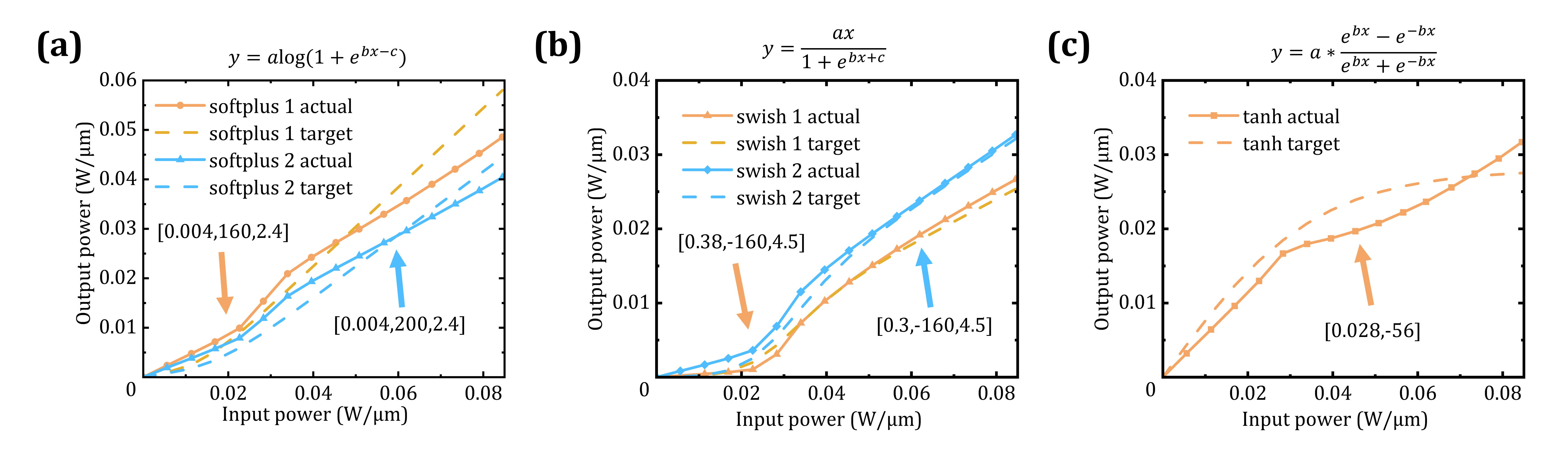}
    \caption{Schematic diagrams of different types of NAFs and partial modifications of NAF shape (a) fitting results of two Softplus-like NAFs with distinct parameters (b) fitting results of two Swish-like NAFs with distinct parameters (c) fitting results of a Tanh-like NAF}
    \label{fig:4}
\end{figure*}

After obtaining the transmition curve of the PA through full-wave simulation, we proceed to acquire diverse reconfigurable NAFs by changing the phase shifter values in the RoNo. In the RoNo, there are three phase shifters whose phase offset values are denoted as $\varphi_1$, $\varphi_2$ and $\varphi_3$. By adjusting these three phase shifters, we can further synthesize the power curves output from the two ports of PA. Suggesting that the y-branch splitter wouldn't bring in phase change and the energy loss in y-branch splitter is neglected. The total output of the RoNo can be calculated using the following formula:
\begin{equation}
     Cons_1 = \frac{1}{2}f_{t}(e_{in})\cos^2\left(\frac{\varphi_2}{2}\right) \label{equ:7}
\end{equation}
\begin{equation}
    Cons_2 = \frac{1}{2}f_{r} (e_{in})\cos^2\left(\frac{\varphi_3}{2}\right) 
    \label{equ:8}
\end{equation}
\begin{equation}
    Cross = \sqrt{f_{t}(e_{in})f_{r}(e_{in})}\cos\varphi_1\cos\left(\frac{\varphi_2}{2}\right)\cos\left(\frac{\varphi_3}{2}\right)
    \label{equ:9}
\end{equation}
\begin{equation}
    I_{total}(e_{in}) = Cons_1 + Cons_2 +Cross
    \label{equ:10}
\end{equation}
Through the \autoref{equ:7}, \autoref{equ:8}, \autoref{equ:9}, \autoref{equ:10}, we can establish a forward model of the NAF synthesis. An ideal NAF was selected and rescaled into the operational energy range of RoNo. The model containing three degrees of freedom is optimized using evolutionary algorithms (e.g., genetic algorithm), with all three variables distributed within the interval $(0,2\pi]$. Three types of NAFs are selected, including softplus, swish and tanh. Furthermore, parameter-specific fine-tuning is performed on both softplus and swish NAFs to validate the model's fine-tuning capability. As shown in \autoref{tab:1}, we employ several scaling parameters to represent the modification of NAF. The first column of the \autoref{tab:1} lists the parametric NAF expressions, the second column specifies the values assigned to the scaling parameters, the following three columns provide the phase shift values generated by the corresponding phase shifters, and the final column indicates the coefficient of determination ($R^2$) between the experimentally measured NAF and our designed NAF, where $R^2>0.9$ demonstrates strong agreement across all configurations. \autoref{fig:4} presents the transmission curves of RoNo under specific phase shifter configurations and their corresponding target NAFs. \autoref{fig:4} (a) and \autoref{fig:4} (b) show the fitting results for the Softplus-like and Swish-like NAFs alongside their parameter-tuned counterparts, respectively. The parametric NAF expressions are displayed above each subfigure, with the corresponding parameter configurations (e.g., [a,b,c]) indicated in parentheses within the \autoref{fig:4}. RoNo demonstrated robust fitting capabilities for both functionally distict NAF and parameter-tuned configurations, achieving high $R^2$ values acrossall test cases, which quantitatively confirms its precision in modeling complex neural activation dynamics.

\begin{table*}[h!]
   \caption{Synthesis of several NAFs\label{tab:1}}
   \centering
   \small 
   \renewcommand{\arraystretch}{1.5}
   \begin{tabularx}{0.7\textwidth}{>{\raggedright}X c *{3}{S[table-format=1.3]} S[table-format=1.4]}
    \toprule
    \textbf{NAF type} & \textbf{parameters} & {\boldmath$\varphi_1$ (rad)} & {\boldmath$\varphi_2$ (rad)} & {\boldmath$\varphi_3$ (rad)} & {\boldmath$R^2$} \\
    \midrule
    $y = a\log(1 + e^{bx - c})$ 
    & {[}0.004, 160, 2.4{]} & 0      & 0      & 1.571 & 0.9598 \\
    
    $y = a\log(1 + e^{bx - c})$ 
    & {[}0.004, 200, 2.4{]} & 0.942  & 0.451  & 1.571 & 0.9573 \\
    
    $\displaystyle y = \frac{ax}{1 + e^{bx + c}}$ 
    & {[}0.3, -160, 4.5{]}  & 0.314  & 0.795  & 3.141 & 0.9937 \\
    
    $\displaystyle y = \frac{ax}{1 + e^{bx + c}}$ 
    & {[}0.38, -160, 4.5{]} & 0.314  & 0.823  & 2.498 & 0.9876 \\
    
    $y = a \tanh(bx)$ 
    & {[}0.028, -56{]}      & 0.314  & 1.617  & 0     & 0.9011 \\
    \bottomrule
  \end{tabularx}
\end{table*}

\section{Application on tasks}

To systematically evaluate RoNo-generated NAFs, we develop two different neural networks benchmarking framework comprising distinct architectural complexities. In both neural networks, conventional NAFs are sustituted with RoNo-generated NAFs, accompanied by signal conditional protocols to eliminate scaling bias between the electrical operating range and traning dataset dimensions.  These two neural networks with different architectures are subsequently deployed to establish a multi-domain benchmarking framework, where the simplified CNN addressed MNIST handwritten digit recognition (10-class, $28\times 28$ graysclae) while the enhanced configuration modified ResNet-18 tackled CIFAR-10 recognition challenges (10-class, $32\times 32$ RGB). All computational experiments are implemented in Pytorch 2.0.0 with architecture-specific optimizations, batch size=32 for MNIST and 16 for CIFAR-10 (batch size=16), executed on NVIDIA RTX3090 GPUs with CUDA 12.4 parallel computing architecture.

\subsection{MNIST with simple CNN}

The MNIST handwritten digit classification task is seen as one of the most versatile dataset for getting started in machine learning, and it contains 60,000 training samples and 10,000 test samples that the network needs to classify recognized pictures into ten categories of numbers ranging from 0 to 9. So, we set up a simple network with only 3 layers to test the effect of the nonlinear activation function we fit. 

\begin{figure}[h!]
    \centering
    \includegraphics[width=0.8\textwidth]{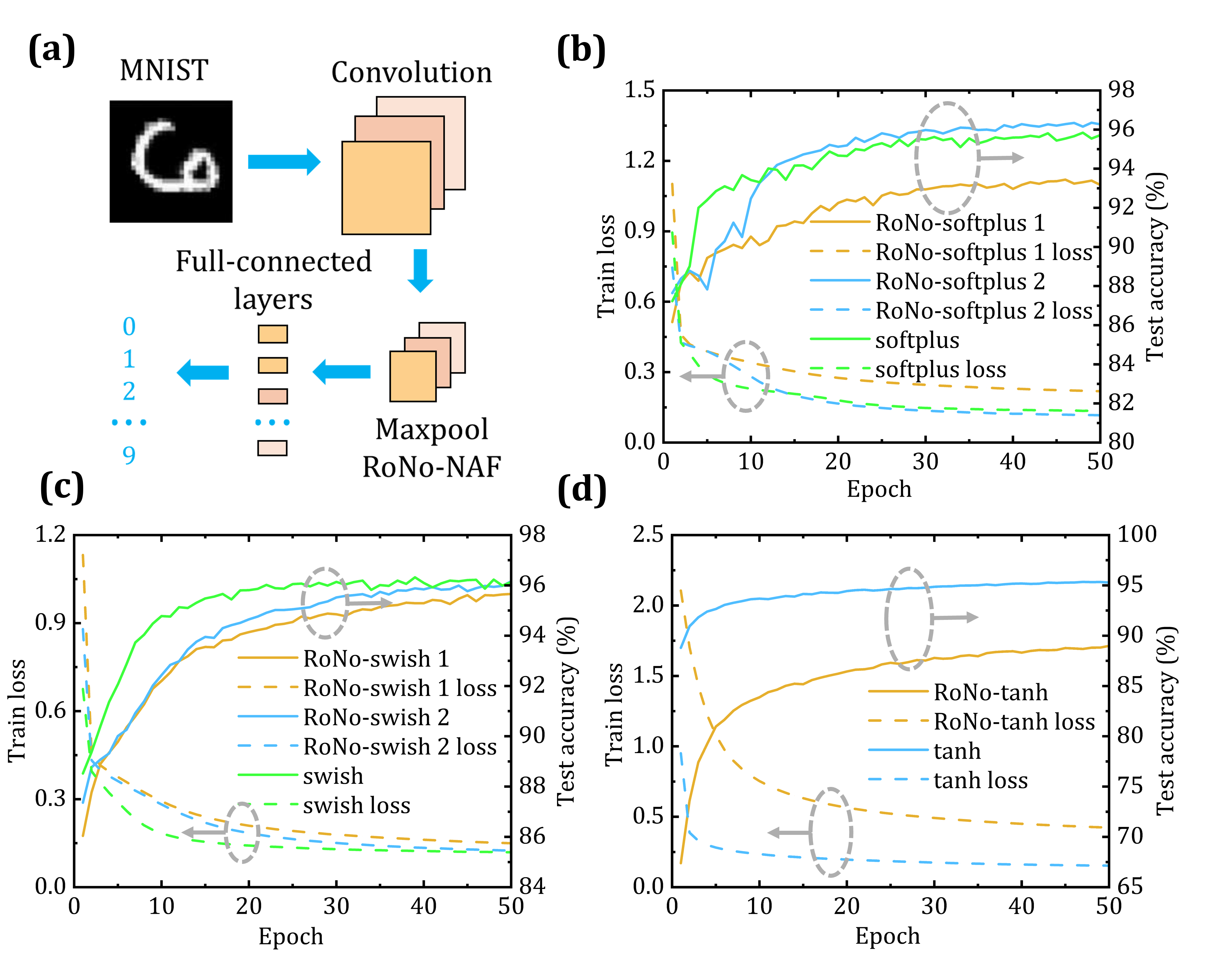}
    \caption{Utilizing a simple CNN using the NAFs generated by RoNo to complete MNIST handwritten task and the test accuracy (solid line) and train loss (dashed line) (a) architecture of the simple CNN (b)~(d) RoNo-generated NAFs versus baseline NAFs in test accuracy and training loss}
    \label{fig:5}
\end{figure}

The architecture of the CNN is shown in \autoref{fig:5} (a). It consists of a single convolutional layer with 4 filters ($3\times 3$ kernel, stride=2) followed by the RoNo-generated NAF and $2\times 2$ max-pooling, progressively reducing spatial dimensions from $28\times 28$ to $7\times 7$ while sxpanding to 4 channels, culminating in a fully-connected layer that maps the 196-dimensional flattened features ($4\times 7 \times 7$) to 10-class outputs, intentionally omitting batch normalization and intermediate dense layers to maintain parameter efficiency (2,010 trainable trainable parameters) while isolating NAF impacts. The rest three figures demonstrates the test accuracy and training loss curves for different RoNo-generated NAFs after 50 training epochs. \autoref{fig:5} (b) comparatively evaluates three Softplus NAFs (detail of results listed in \autoref{tab:2}): the parametrically tuned RoNo-Softplus1 ( $93.44\%$ test accuracy / 0.218 training loss) and RoNo-Softplus2 ($96.34\%$ / 0.114) partially surpass the baseline Softplus ($95.72\%$ / 0.135). \autoref{fig:5} (c) demonstrates that two tuned Swish variants: RoNo-Swish1 ($95.67\%$ / 0.149) and RoNo-Swish2 ($96.16\%$ / 0.124), remaining competitive against the base line Swish ($95.96\%$ / 0.149), with complete results documented in \autoref{tab:2}. As is shown in \autoref{fig:5} (d) and numerically listed in \autoref{tab:2}, the RoNo-Tanh ($89.12\%$ / 0.421) underperforms its baseline couterpart ($95.29\%$ / 0.153).
The systematic comparisons in \autoref{fig:5} (supported by metrics in \autoref{tab:2}) demonstrate that our RoNo can generate high-quality NAFs that exhibit excellent perfoermance in shallow networks and simple tasks. Meanwhile, the diverse NAFs generated by our RoNo can closely approximate or even surpass their standard counterparts in electrical environments, further validating the superior performance of our framework. In the following, we would conduct further comprehensive validation of NAFs generated by RoNo in deep neural networks and more complex tasks.

\begin{table}[h!]
  \centering
  \caption{Experimental Results of MNIST}
  \renewcommand{\arraystretch}{1.2}
  \begin{tabularx}{0.45\textwidth}{l S[table-format=2.2] S[table-format=1.3]} 
    \toprule
    \textbf{NAF type} & \textbf{test accuracy (\%)} & \textbf{train loss} \\
    \midrule
    RoNo-Softplus1 & 93.44 & 0.218 \\
    RoNo-Softplus2 & 96.34 & 0.114 \\
    RoNo-Swish1     & 95.67 & 0.149 \\
    RoNo-Swish2     & 96.16 & 0.124 \\
    RoNo-Tanh       & 89.12 & 0.421 \\
    Swish           & 95.96 & 0.119 \\
    Softplus        & 95.72 & 0.135 \\
    Tanh            & 95.29 & 0.153 \\
    \bottomrule
  \end{tabularx}
  \label{tab:2}
\end{table}


\subsection{CIFAR-10 with modified ResNet}

The CIFAR-10 image classification benchmark is a widely adopted dataset for evaluating computer vision models, consisting of 50,000 training and 10,000 testing RGB images divided into ten object categories such as airplanes, automobiles, birds, and cats. These $32\times 32$-pixel images pose significant challenges due to their low resolution, color diversity, and complex real-world object features. To rigorously assess the impact of activation function customization in deep architectures, we developed a modified ResNet-18 framework \cite{ResNet} where standard NAF (e.g., ReLU) are comprehensively replaced with the RoNo-NAF across all convolutional and fully connected layers. This systematic replacement methodology enables precise evaluation of how RoNo-NAF influences the final classification performance in resource-constrained but semantically rich visual recognition tasks.

\begin{figure}[h!]
    \centering
    \includegraphics[width=0.8\textwidth]{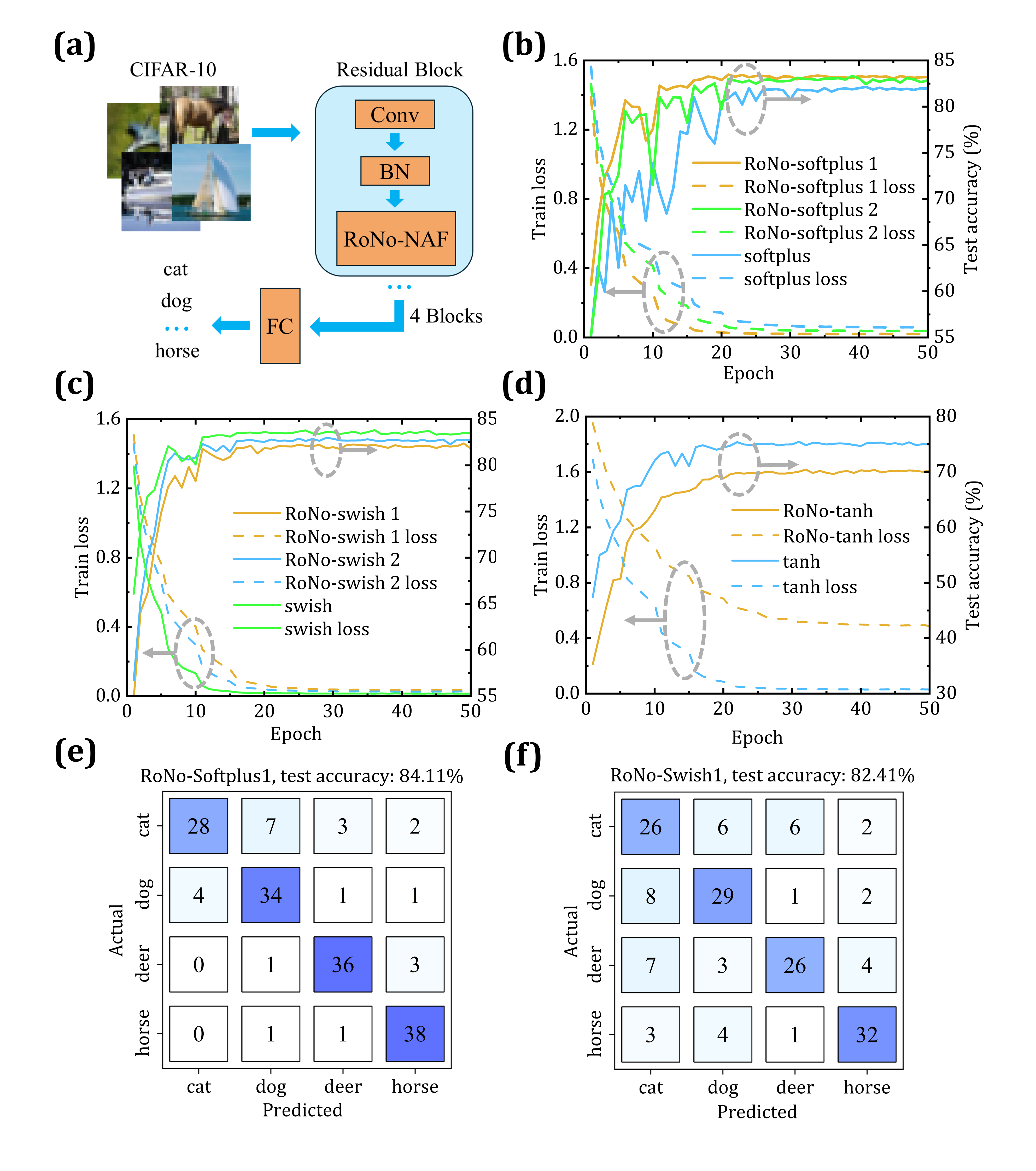}
    \caption{Utilizing modified ResNet-18 using the NAFs generated by RoNo to complete MNIST handwritten task (a) architecture of the modified ResNet-18 (b)~(d) RoNo-generated NAFs versus baseline NAFs in test accuracy and training loss (e),(f) confusion matrix of the modified ResNet-18 with RoNo-Softplus and RoNo-Swish}
    \label{fig:6}
\end{figure}

\autoref{fig:6} (a) illustrates the modified ResNet-18 architecture where all standard NAFs in residual blocks are systematically replaced with RoNo-generated NAFs, while retaining the original convolutional layers, skip connections and layer dimensions to preserve the network's intrinsic topology. This controlled substitution method isolates the impact of photonic NAFs on network performance without introducing structural confounding factors. The remaining five figures demonstrate the test accuracy, training loss and representative confusion matrices of various RoNo-generated NAFs after 50 training epochs (detail of results listed in \autoref{tab:3}). \autoref{fig:6} (b) comparatively analyzes three Softplus variants: RoNo-Softplus1 ($83.32\%$ test accuracy / 0.021 training loss) and RoNo-Softplus2 ($82.84\%$ / 0.038), both outperforming the baseline Softplus NAF ($81.96\%$ / 0.058). \autoref{fig:6} (c) show the comparative performance of three Swish variants: RoNo-Swish1 ($82.09\%$ / 0.037), RoNo-Swish2 ($82.78\%$ / 0.027), and the baseline Swish ($83.52\%$ / 0.015). \autoref{fig:6} (d) demonstrates the comparative performance between the RoNo-Tanh ($70.22\%$ / 0.488) and the baseline Tanh ($74.87\%$ / 0.031). \autoref{fig:6} (e) and \autoref{fig:6} (f) reveal key performance difference: RoNo-Softplus1 ($84.11\%$ test accuracy) reduces cross-species misclassifications compared to RoNo-Swish1 ($82.41\%$), particularly in deer recognition (36 vs 26 correct predictions) and horse classification (38 vs 32). While both models struggle most with cat-dog confusion, RoNo-Softplus1 halves cat to dog errors (7 to 3) and eliminates deer to horse misclassification (4 to 0), demonstrating optimized feature discrimination through parametric tuning of RoNo-generated NAFs.

The systematic comparisons in \autoref{fig:6} demonstrate that RoNo-generated on-chip NAF can achieve competitive performance in deep neural networks handling complex tasks, closely matching or even exceeding conventional electrical NAF implementations. Nevertheless, specific NAF profiles (e.g., Tanh variants) exhibit convergence challenges during training. These experimental results highlight the necessity of real-time reconfigurable NAF in broader all-optical on-chip training scenarios to address various types of networks and tasks.
\begin{table}[ht]
  \centering
  \caption{Experimental Results of CIFAR-10}
  \renewcommand{\arraystretch}{1.2}
  \begin{tabularx}{0.45\textwidth}{l S[table-format=2.2] S[table-format=1.3]} 
    \toprule
    \textbf{NAF type} & \textbf{test accuracy (\%)} & \textbf{train loss} \\
    \midrule
    RoNo-Softplus1 & 84.11 & 0.021 \\
    RoNo-Softplus2 & 82.84 & 0.038 \\
    RoNo-Swish1     & 82.41 & 0.037 \\
    RoNo-Swish2     & 82.78 & 0.027 \\
    RoNo-Tanh       & 70.22 & 0.488 \\
    Softplus        & 81.96 & 0.058 \\
    Swish           & 83.52 & 0.015 \\
    Tanh            & 74.87 & 0.031 \\
    \bottomrule
  \end{tabularx}
  \label{tab:3}
\end{table}
\section{Conclusion}
In this work, we experimentally demonstrate a chip-integrated all-optical reconfigurable nonlinear activation generator (RoNo) serving as a nonlinear activation function (NAF) generator in neural computing systems. The results empirically validate that RoNo enables real-time reconfiguration of optical NAFs approximating arbitrary mathematical forms, achieving superior performance in both convolutional and residual network architectures across classification and feature learning tasks. Through coherent photonic computation principles, our optical NAF implementations deliver comparable or enhanced classification efficacy relative to electronic counterparts while fundamentally bypassing optoelectronic interface constraints. This advancement establishes a critical pathway toward scalable optical neural processors and energy-efficient photonic computing paradigms.

\section*{Acknowledgment}

Funding: National Natural Science Foundation of China (62171055, 62135009, 62471062); Fundamental Research
Funds for the Central Universities (ZDYY202102); BUPT Innovation and Entrepreneurship Support Program (2021-YCA111), P. R. China; Super Computing Platform of Beijing University of Posts and Telecommunications.

\bibliographystyle{ieeetr}
\bibliography{Ref}

\begin{thebibliography}{10}

\bibitem{chen2025deep}
X.~Chen, X.~Hu, Y.~Huang, H.~Jiang, W.~Ji, Y.~Jiang, Y.~Jiang, B.~Liu, H.~Liu, X.~Li, {\em et~al.}, ``Deep learning-based software engineering: progress, challenges, and opportunities,'' {\em Science China Information Sciences}, vol.~68, no.~1, pp.~1--88, 2025.

\bibitem{deeplearning_genomic}
M.~Habibur, M.~A. Rahman, M.~A. Bakar, M.~Mostafizur, M.~Afroza, {\em et~al.}, ``Significance of artificial intelligence in clinical and genomic diagnostics,'' {\em Journal of Precision Biosciences}, vol.~7, no.~1, pp.~1--14, 2025.

\bibitem{deeplearning_material}
X.~Zhong, B.~Gallagher, S.~Liu, B.~Kailkhura, A.~Hiszpanski, and T.~Y.-J. Han, ``Explainable machine learning in materials science,'' {\em npj computational materials}, vol.~8, no.~1, p.~204, 2022.

\bibitem{deeplearning_image}
E.~Maggiori, Y.~Tarabalka, G.~Charpiat, and P.~Alliez, ``Convolutional neural networks for large-scale remote-sensing image classification,'' {\em IEEE Transactions on geoscience and remote sensing}, vol.~55, no.~2, pp.~645--657, 2016.

\bibitem{deeplearning_photonic1}
W.~Ma, Z.~Liu, Z.~A. Kudyshev, A.~Boltasseva, W.~Cai, and Y.~Liu, ``Deep learning for the design of photonic structures,'' {\em Nature photonics}, vol.~15, no.~2, pp.~77--90, 2021.

\bibitem{deeplearning_photonic2}
C.~Zuo, J.~Qian, S.~Feng, W.~Yin, Y.~Li, P.~Fan, J.~Han, K.~Qian, and Q.~Chen, ``Deep learning in optical metrology: a review,'' {\em Light: Science \& Applications}, vol.~11, no.~1, pp.~1--54, 2022.

\bibitem{deeplearning_photonic3}
H.~P. Wang, Y.~B. Li, H.~Li, S.~Y. Dong, C.~Liu, S.~Jin, and T.~J. Cui, ``Deep learning designs of anisotropic metasurfaces in ultrawideband based on generative adversarial networks,'' {\em Advanced Intelligent Systems}, vol.~2, no.~9, p.~2000068, 2020.

\bibitem{deeplearning_photonic4}
L.~Bian, X.~Zhan, R.~Yan, X.~Chang, H.~Huang, and J.~Zhang, ``Physical twinning for joint encoding-decoding optimization in computational optics: a review,'' {\em Light: Science \& Applications}, vol.~14, no.~1, p.~162, 2025.

\bibitem{ONN1}
R.~Hamerly, L.~Bernstein, A.~Sludds, M.~Solja{\v{c}}i{\'c}, and D.~Englund, ``Large-scale optical neural networks based on photoelectric multiplication,'' {\em Physical Review X}, vol.~9, no.~2, p.~021032, 2019.

\bibitem{ONN2}
Y.~Huang, T.~Fu, H.~Huang, S.~Yang, and H.~Chen, ``Sophisticated deep learning with on-chip optical diffractive tensor processing,'' {\em Photon. Res.}, vol.~11, pp.~1125--1138, Jun 2023.

\bibitem{NAF1}
H.~Chen, Z.~Yu, T.~Zhang, Y.~Zang, Y.~Dan, and K.~Xu, ``Advances and challenges of optical neural networks,'' {\em Chin. J. Lasers}, vol.~47, no.~05, pp.~80--91, 2020.

\bibitem{NAF2}
T.~Szanda{\l}a, ``Review and comparison of commonly used activation functions for deep neural networks,'' in {\em Bio-inspired neurocomputing}, pp.~203--224, Springer, 2020.

\bibitem{zili_cai_inverse_design}
Z.~Cai, Z.~Yang, T.~Zhang, J.~Dai, K.~Xu, and Q.~Chen, ``A realization of activation function based on inverse design,'' in {\em 2024 IEEE Opto-Electronics and Communications Conference (OECC)}, pp.~1--3, 2024.

\bibitem{electric_optical_NAF1}
M.~M. Pour~Fard, I.~A. Williamson, M.~Edwards, K.~Liu, S.~Pai, B.~Bartlett, M.~Minkov, T.~W. Hughes, S.~Fan, and T.-A. Nguyen, ``Experimental realization of arbitrary activation functions for optical neural networks,'' {\em Optics Express}, vol.~28, no.~8, pp.~12138--12148, 2020.

\bibitem{electric_optical_NAF2}
I.~A. Williamson, T.~W. Hughes, M.~Minkov, B.~Bartlett, S.~Pai, and S.~Fan, ``Reprogrammable electro-optic nonlinear activation functions for optical neural networks,'' {\em IEEE Journal of Selected Topics in Quantum Electronics}, vol.~26, no.~1, pp.~1--12, 2019.

\bibitem{saturable_absorption_NAF1}
H.~Tari, A.~Bile, F.~Moratti, and E.~Fazio, ``Sigmoid type neuromorphic activation function based on saturable absorption behavior of graphene/pmma composite for intensity modulation of surface plasmon polariton signals,'' {\em Plasmonics}, vol.~17, no.~3, pp.~1025--1032, 2022.

\bibitem{saturable_absorption_NAF2}
K.~Liao, C.~Li, T.~Dai, C.~Zhong, H.~Lin, X.~Hu, and Q.~Gong, ``Matrix eigenvalue solver based on reconfigurable photonic neural network,'' {\em Nanophotonics}, vol.~11, no.~17, pp.~4089--4099, 2022.

\bibitem{PCM_MRR_NAF}
T.~Y. Teo, X.~Ma, E.~Pastor, H.~Wang, J.~K. George, J.~K. Yang, S.~Wall, M.~Miscuglio, R.~E. Simpson, and V.~J. Sorger, ``Programmable chalcogenide-based all-optical deep neural networks,'' {\em Nanophotonics}, vol.~11, no.~17, pp.~4073--4088, 2022.

\bibitem{germanium-silicon_NAF}
Y.~Shi, J.~Ren, G.~Chen, W.~Liu, C.~Jin, X.~Guo, Y.~Yu, and X.~Zhang, ``Nonlinear germanium-silicon photodiode for activation and monitoring in photonic neuromorphic networks,'' {\em Nature Communications}, vol.~13, no.~1, p.~6048, 2022.

\bibitem{NAF_comparison}
D.~Gangadia, ``Activation functions: Experimentation and comparison,'' in {\em 2021 6th International Conference for Convergence in Technology (I2CT)}, pp.~1--6, IEEE, 2021.

\bibitem{HuangChaoRan_reconfigurable}
A.~Jha, C.~Huang, and P.~R. Prucnal, ``Reconfigurable all-optical nonlinear activation functions for neuromorphic photonics,'' {\em Optics letters}, vol.~45, no.~17, pp.~4819--4822, 2020.

\bibitem{inverse_design_NSGA}
Y.~Dan, Z.~Fan, X.~Sun, T.~Zhang, and K.~Xu, ``All-type optical logic gates using plasmonic coding metamaterials and multi-objective optimization,'' {\em Optics Express}, vol.~30, no.~7, pp.~11633--11646, 2022.

\bibitem{inverse_design_deeplearning1}
P.~R. Wiecha, A.~Arbouet, C.~Girard, and O.~L. Muskens, ``Deep learning in nano-photonics: inverse design and beyond,'' {\em Photonics Research}, vol.~9, no.~5, pp.~B182--B200, 2021.

\bibitem{inverse_design_deeplearning2}
S.~Yu, T.~Zhang, J.~Dai, and K.~Xu, ``Hybrid inverse design scheme for nanophotonic devices based on encoder-aided unsupervised and supervised learning,'' {\em OPTICS EXPRESS}, vol.~31, pp.~39852--39866, NOV 20 2023.

\bibitem{adjoint_method_and_inverse_design_for_nonlinear_nanophotonic_devices}
T.~W. Hughes, M.~Minkov, I.~A.~D. Williamson, and S.~Fan, ``Adjoint method and inverse design for nonlinear nanophotonic devices,'' {\em ACS PHOTONICS}, vol.~5, pp.~4781--4787, DEC 2018.

\bibitem{Reconfigurable_nonlinear_optical_element_using_tunable_couplers_and_inverse-designed_structure}
V.~Nikkhah, M.~J. Mencagli, and N.~Engheta, ``Reconfigurable nonlinear optical element using tunable couplers and inverse-designed structure,'' {\em NANOPHOTONICS}, vol.~12, pp.~3019--3027, JUL 14 2023.

\bibitem{Measurement_of_the_Kerr_Nonlinear_Refractive_Index_and_its_Variation_Among_4H-SiC_Wafers}
J.~Li, R.~Wang, L.~Cai, and Q.~Li, ``Measurement of the kerr nonlinear refractive index and its variation among 4h-sic wafers,'' {\em PHYSICAL REVIEW APPLIED}, vol.~19, MAR 24 2023.

\bibitem{Initial_value_for_inverse_design}
Z.~Wang, B.-Z. Wang, J.-P. Liu, and R.~Wang, ``Method to obtain the initial value for the inverse design in nanophotonics based on a time-reversal technique,'' {\em OPTICS LETTERS}, vol.~46, pp.~2815--2818, JUN 15 2021.

\bibitem{Gauss-Newton_and_full_Newton_methods_in_frequency-space_seismic_waveform_inversion}
R.~Pratt, C.~Shin, and G.~Hicks, ``Gauss-newton and full newton methods in frequency-space seismic waveform inversion,'' {\em GEOPHYSICAL JOURNAL INTERNATIONAL}, vol.~133, pp.~341--362, MAY 1998.

\bibitem{ResNet}
K.~He, X.~Zhang, S.~Ren, and J.~Sun, ``Deep residual learning for image recognition,'' in {\em 2016 IEEE CONFERENCE ON COMPUTER VISION AND PATTERN RECOGNITION (CVPR)}, IEEE Conference on Computer Vision and Pattern Recognition, pp.~770--778, IEEE Comp Soc; Comp Vis Fdn, 2016.
\newblock 2016 IEEE Conference on Computer Vision and Pattern Recognition (CVPR), Seattle, WA, JUN 27-30, 2016.

\end{thebibliography}

\end{document}